\documentclass{webofc}
\usepackage[varg]{txfonts}   
\usepackage{ptdr-definitions}
\usepackage{hepunits}
\usepackage{heppennames2}
\usepackage{amsmath}
\usepackage{amssymb}
\usepackage{xpatch}
\makeatletter
\xpatchcmd\@HepConStyle
 {\edef\@upcode{\updefault}}
 {\ifdefined\shapedefault\edef\@upcode{\shapedefault}\else\edef\@upcode{\updefault}\fi}
 {}{}
\makeatother
\RequirePackage[colorlinks,citecolor=blue,urlcolor=blue,linkcolor=blue]{hyperref}
\usepackage{adjustbox}
\usepackage{multirow}

\newcommand{\PbPb}{\ensuremath{\mathrm{Pb}\mathrm{Pb}}\xspace}
\newcommand{\pp}{\ensuremath{\Pp\Pp}\xspace}

\newcommand{\RAA}{$R_\mathrm{AA}$\xspace}

\begin{document}

\title{Charm quark and QGP interactions through the spectra and anisotropic flow of D$^0$ over the widest p$_\text{T}$ interval using event-shape engineering at CMS}

\author{\firstname{Soumik} \lastname{Chandra}\inst{1}\fnsep\thanks{\email{chand140@purdue.edu}} (on behalf of the CMS Collaboration)}

\institute{Purdue University, West Lafayette, IN, USA}

\abstract{The charm quark is formed almost exclusively during the initial stages of the collision, and a significant fraction of the charm quarks fragment into the $\mathrm{D}^{0}$ meson, the lightest open-charm hadron. We can gain insights into the interactions between the charm quark and the quark-gluon plasma (QGP) medium by studying the production and the flow of the $\mathrm{D}^{0}$ meson in heavy-ion collisions. We study the effect of the initial shape of the collision system on the elliptic flow ($v_{2}$) of promptly produced $\mathrm{D}^{0}$ mesons in the $p_\mathrm{T}$ region 2--30\GeVc using event-shape engineering (ESE) in PbPb collisions at 5.02 TeV at the CMS experiment. A correlation between the initial shape anisotropy and the $\mathrm{D}^{0}$ $v_{2}$ would suggest that the flow gets driven by the interactions between the charm quark and the QGP. Comparison with theoretical predictions allows us to unravel the mechanism behind the generation of $v_{2}$. We also study the anisotropic flow of nonprompt $\mathrm{D}^{0}$ produced due to the decay of the bottom quarks and the nuclear modification factor ($\mathrm{R_{AA}}$) for prompt and nonprompt $\mathrm{D}^{0}$. These studies provide further insights into the QGP interactions of heavy quarks like charm and bottom in heavy-ion collisions in different $p_\mathrm{T}$ regions ranging from 1--40 GeV/c, the widest ever performed with ESE, and centralities between 0--50\%.}
\maketitle
\section{Introduction}
\label{introduction}
The heavy ions colliding at high energies produce a deconfined state of quarks and gluons, known as the Quark-Gluon Plasma (QGP)~\cite{QGP, QGP2, QGP3}. Charm quarks are produced dominantly at the early stages of collision due to their large rest mass; they follow the medium evolution, and the study of charm quarks provides insights into different methods of hadronization in the QGP. The energy loss of charm quarks traveling through the QGP is interesting since quarks can lose energy via elastic and inelastic interactions with the medium~\cite{charm}. The elliptic flow ($v_2$) is the manifestation of the collective flow of particles produced in the collision, and the fluctuations in the initial shape drive the triangular flow ($v_3$). The initial eccentricity of the collision system translates into the elliptic flow of charged hadrons at low \pt~\cite{low_pt_v2}. Studying the elliptic flow of charmed hadrons, such as \PDz, offers a crucial perspective into the energy-loss mechanism and the thermalization of charm quarks in the QGP medium. We also measure the nuclear modification factor (\RAA) to study the suppression of the charm hadrons produced in large systems (like \PbPb collisions) compared to smaller systems (\pp collisions).

In this contribution, we report on the measurements of promptly produced \PDz meson $v_2$ at the center-of-mass energy per nucleon, $\sqrtsNN=5.02$~TeV in \PbPb collisions collected by the CMS detector in the Run2 era~\cite{JINST, HIN-24-015_PAS}, measured in five centrality classes (0--10, 10--20, 20--30, 30--40, and 40--50\%) and ten classes of the initial shape eccentricity in the mid-rapidity region ($\abs{y}<1$). The low-\pt (1--3\GeVc) charged particle $v_2$ measured in the same centrality and initial-shape eccentricity classes, and in $\abs{\eta}<1$ is correlated with the \PDz $v_2$ in the \pt 2--30\GeVc. We measure the low-\pt (1--3\GeVc) charged particle $v_2$ in the same centrality and initial-shape eccentricity classes, and study the correlation with the \PDz $v_2$. 

We also report the $v_2$, $v_3$, and the \RAA of both promptly produced \PDz and the nonprompt \PDz, produced due to the decay of \PQb quark. The study of the prompt and nonprompt \PDz allows us to understand the properties of charm and bottom quarks~\cite{RAA1, RAA2, HIN-19-008, HIN-21-003}.

\section{Data sets and analysis details}
\label{dataset}
The analyses use \PbPb collision data collected with an integrated luminosity of 0.607~nb$^{-1}$. The \PDz mesons are reconstructed via the decay channel $\PDz\to\PKm\Pgpp(\PaDz\to\PKp\Pgpm)$. Due to the absence of particle identification, the particles are reconstructed using all possible combinations of the decay tracks. A gradient-boosted decision tree (BDTG) algorithm from \textsc{tmva} package is applied to reduce the combinatorial background. The fraction of prompt and nonprompt \PDz is extracted by a template fit of the distance of the closest approach (DCA) between the collision point and the direction of the \PDz momentum vector.

\section{Results}
\label{result}
\begin{figure}[!h]
    \centering
    \includegraphics[width=0.9\linewidth]{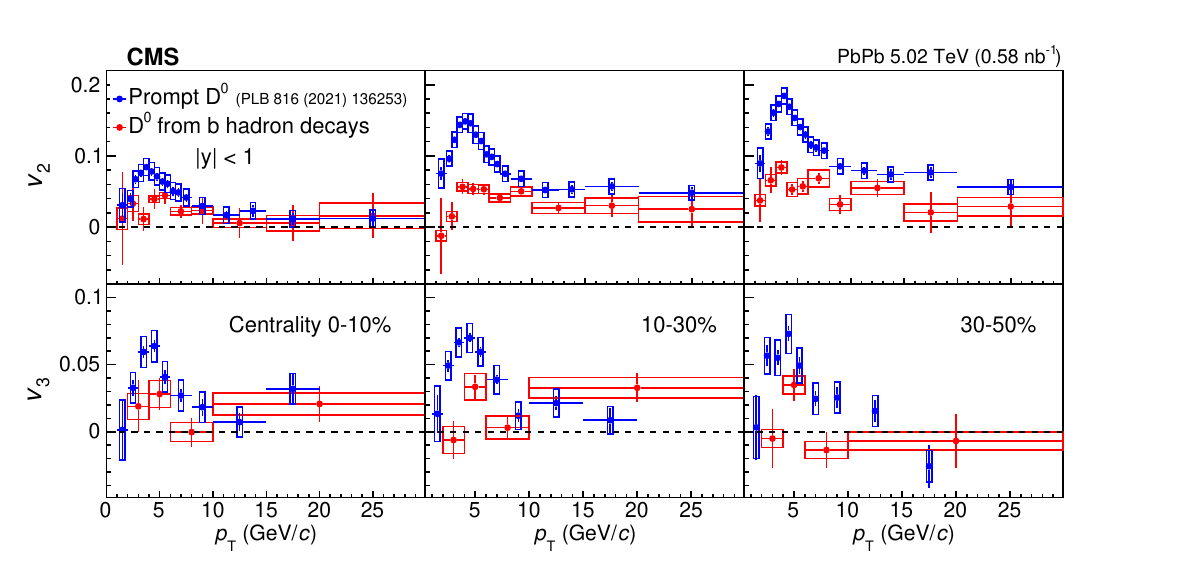}
    \caption{The prompt and nonprompt \PDz $v_2$ (upper panels) and $v_3$ (lower panels) as functions of their \pt and in three bins of centrality. The bars and the boxes represent statistical and systematic uncertainties, respectively. This figure is taken from Ref.~\cite{HIN-21-003}. }
    \label{fig:v2_v3_nonprompt}
\end{figure}

The prompt and nonprompt \PDz \RAA shows the energy loss of the charm quark in the QGP medium, produced in the \PbPb collisions~\cite{RAA1, RAA2}. A strong \pt dependence of the \RAA is observed with the minima around \PDz \pt = 10\GeVc, and the enhancement at low \pt can be attributed to hadronization via coalescence. A strong \pt dependence of the $v_2$ and $v_3$ is observed for prompt \PDz, and the observed $v_3$ is significant for low \pt ($<10\GeVc$)~\cite{HIN-19-008}. The \PDz $v_2$ shows a strong dependence on centrality, while $v_3$ is primarily independent of centrality, as shown in Fig.~\ref{fig:v2_v3_nonprompt}. An indication of a nonzero $v_3$ of nonprompt \PDz with \pt in 4--6\GeVc is observed with a weak dependence on \pt and centrality, and the measured nonprompt \PDz $v_2$ and $v_3$ values are lower than those of the prompt \PDz $v_2$ and $v_3$~\cite{HIN-21-003}.

\begin{figure}[!h]
\centering
\includegraphics[width=0.32\textwidth,clip]{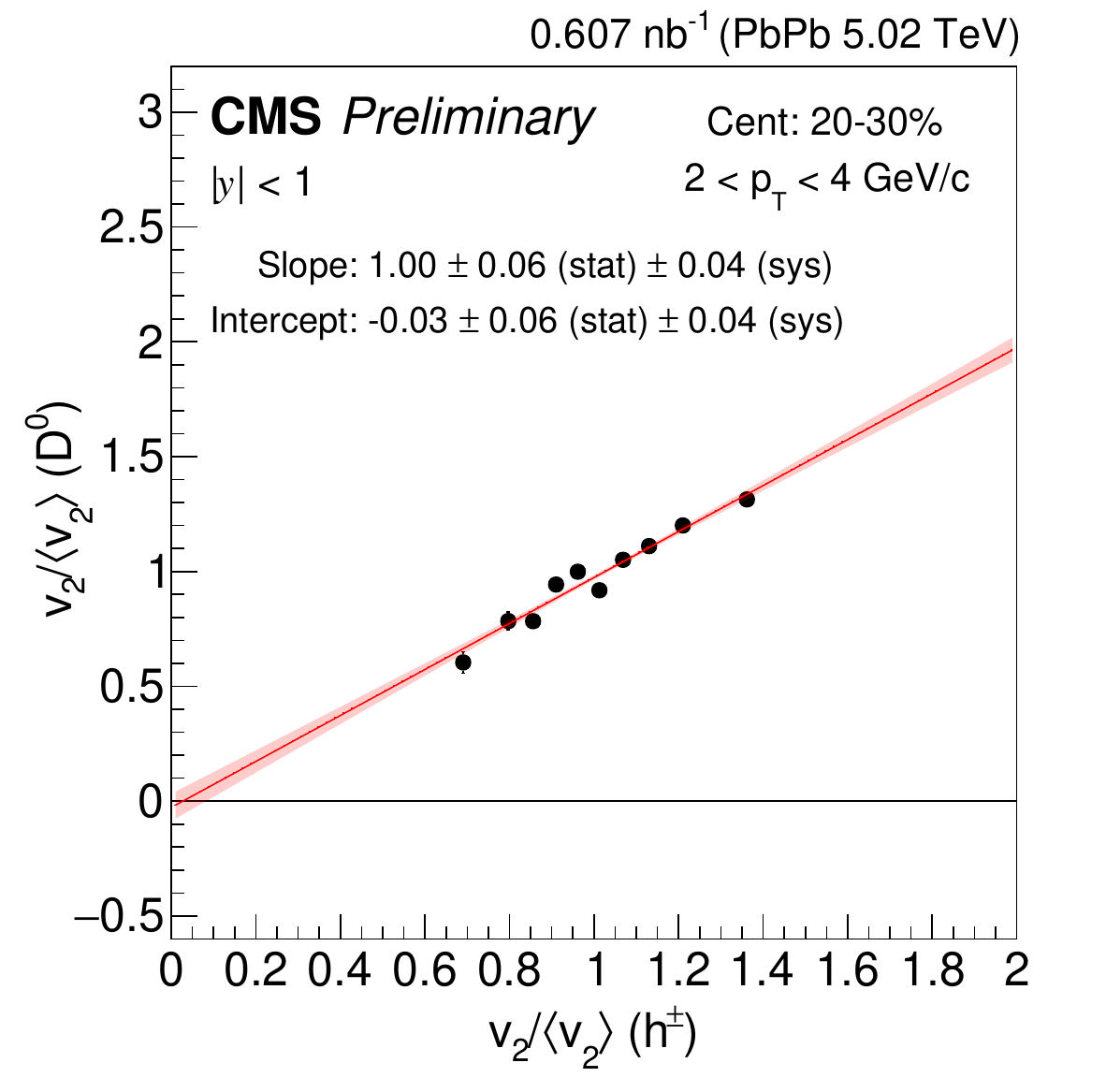}
\includegraphics[width=0.32\textwidth,clip]{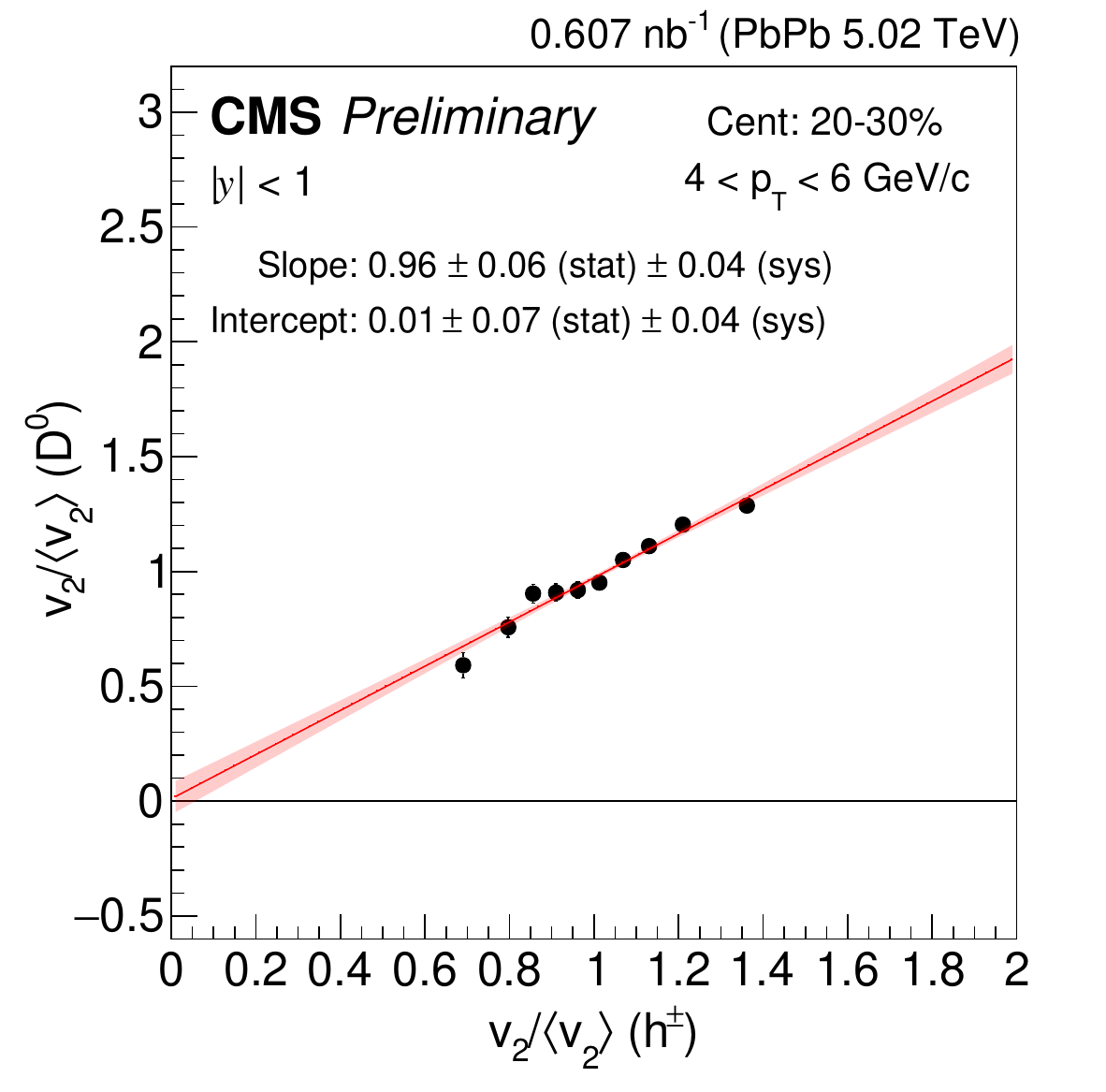}
\includegraphics[width=0.32\textwidth,clip]{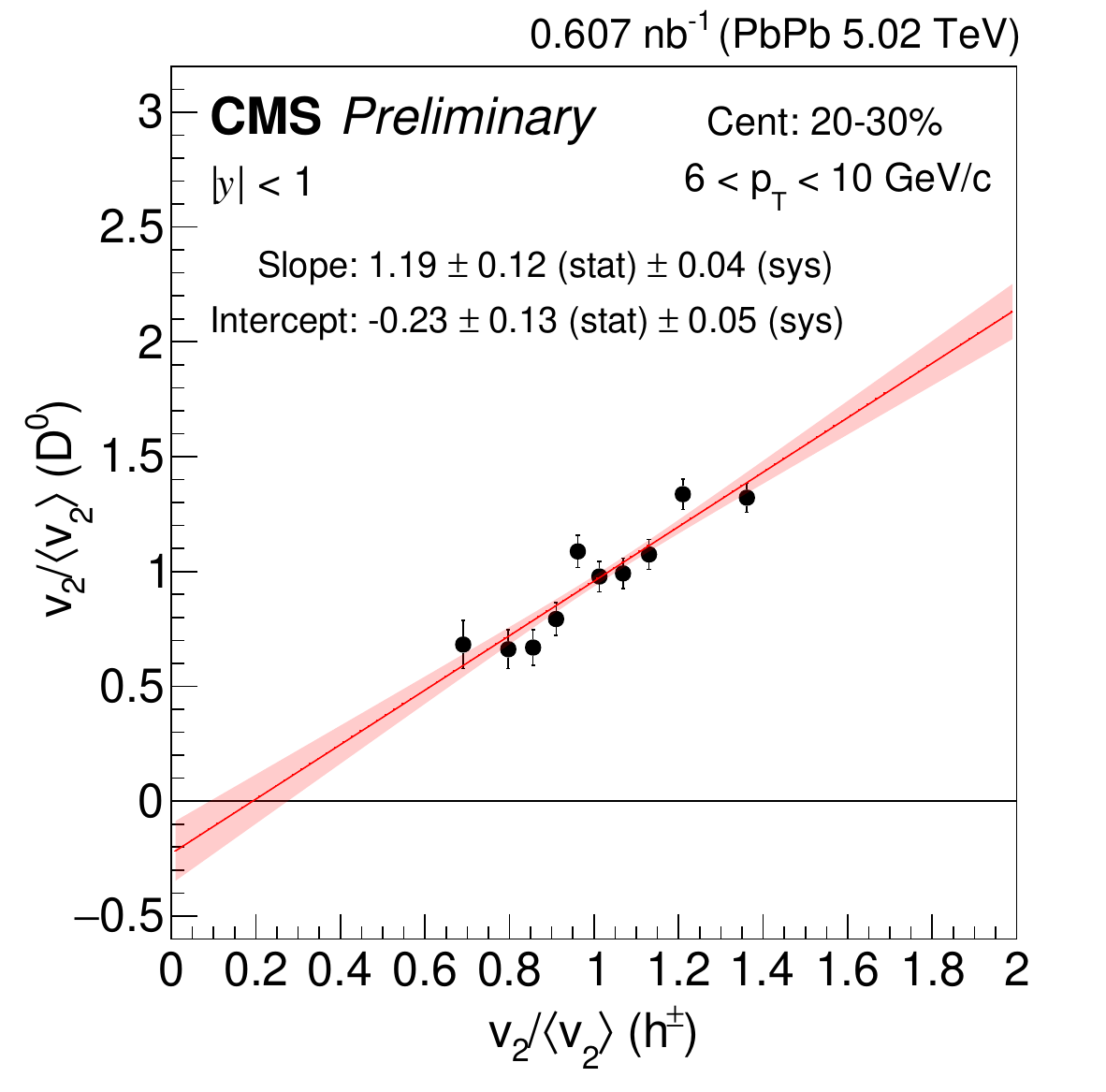}
\caption{Correlation between the normalized \PDz $v_2$ and the charged-particle $v_2$. The $v_2$ value in each $q_2$ bin is scaled by the $v_2$ in each centrality bin. The charged-particle $v_2$ is measured in the \pt region 1--3\GeVc. The correlation plots for \PDz \pt 2--4 (left), 4--6 (middle), and 6--10 (right) in the centrality class 20-30\%. The red band corresponds to the uncertainty of one standard deviation. These figures are taken from Ref.~\cite{HIN-24-015_PAS}}
\label{fig:v2_corr}       
\end{figure}

\begin{figure}[!h]
\centering
\includegraphics[width=0.4\textwidth,clip]{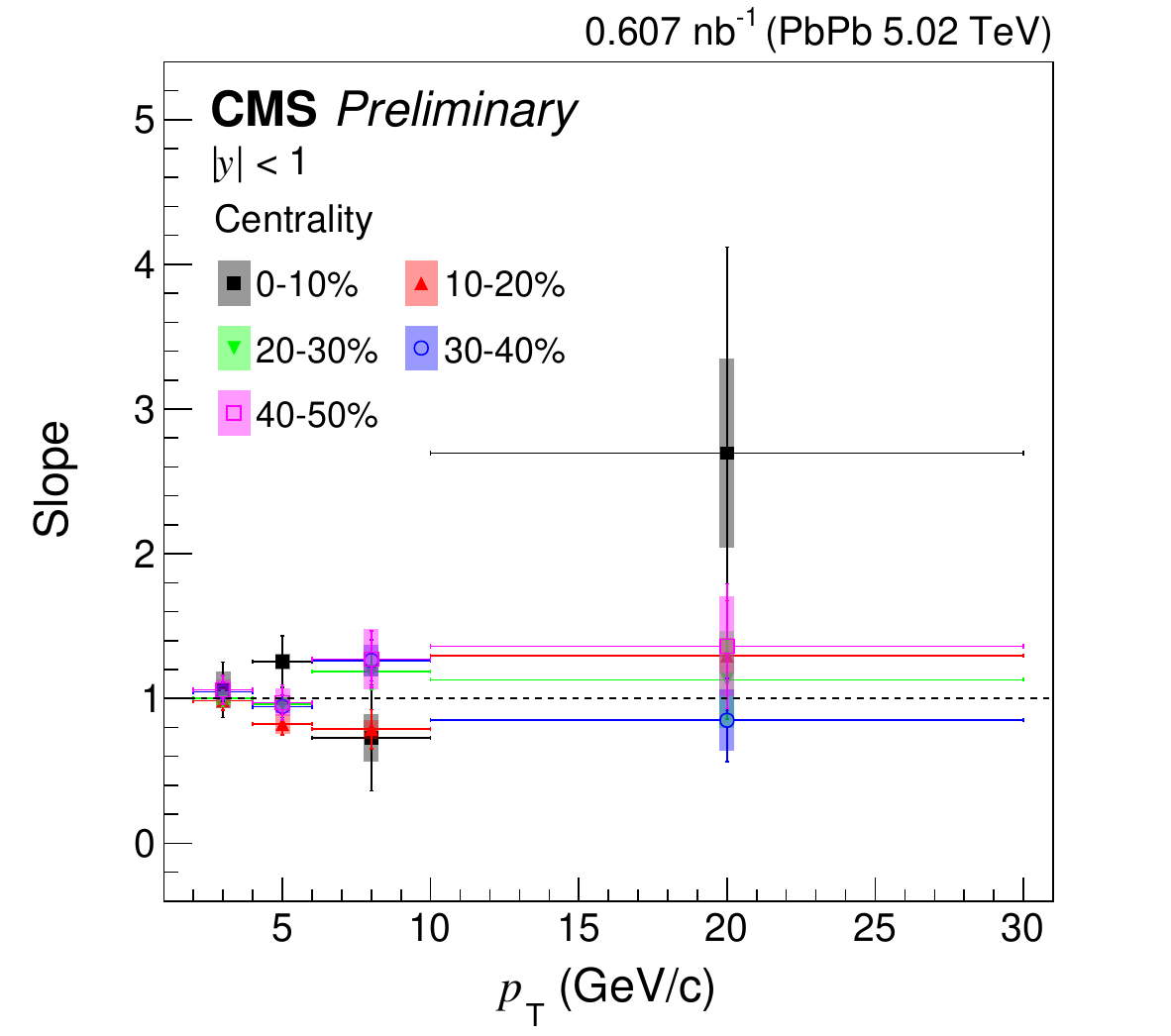}
\includegraphics[width=0.4\textwidth,clip]{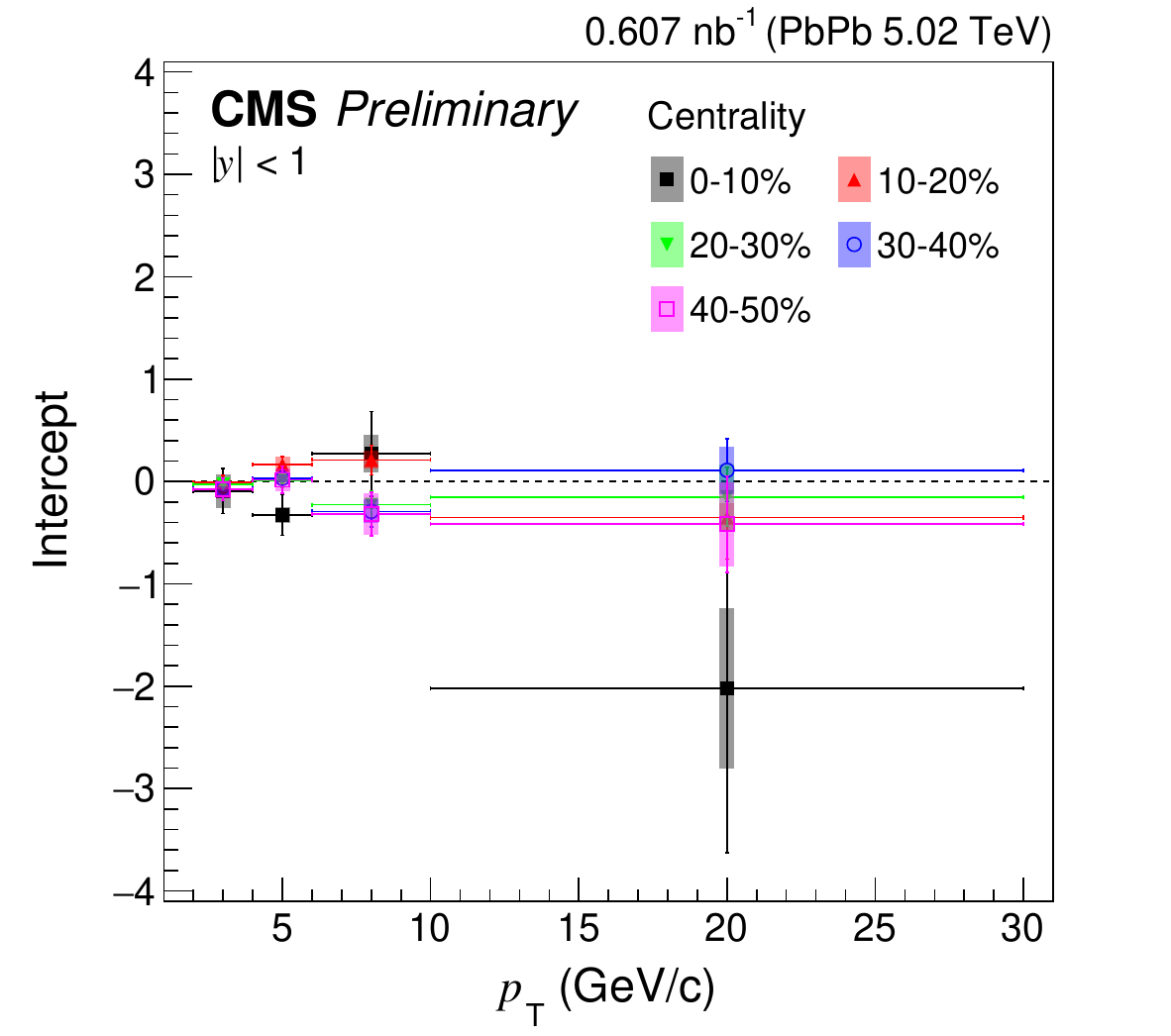}
\caption{The slopes (left) and intercepts (right) obtained from the correlation plots of the prompt \PDz $v_2$ vs the charged-particle $v_2$ for each \pt and centrality class of \PDz. For the correlation plots, the $v_2$ in each $q_2$ bin is normalized by the $q_2$-inclusive $v_2$. The vertical lines correspond to the statistical uncertainties, and the vertical bands correspond to the systematic uncertainties added in quadrature. The slopes and intercepts calculated are consistent with unity and zero, respectively. These figures are taken from Ref.~\cite{HIN-24-015_PAS}}
\label{fig:slope_intercept}       
\end{figure}

The correlation between \PDz $v_2$ and low-\pt (1--3\GeVc) charged-particle $v_2$ is presented in the \PDz \pt region of 2--30\GeVc. The measurement is performed in five centrality classes between 0--50\%, and ten classes of initial shape eccentricity ($q_2$). Fig.~\ref{fig:v2_corr} illustrates the correlation between \PDz and charged-particle $v_2$ for selected \pt and centrality bins. The \pt and centrality dependencies are mitigated by normalizing the \PDz $v_2$ and the charged-particle $v_2$ by their respective elliptic flow within each centrality and \pt bin. The normalized \PDz and charged-particle $v_2$ correlation exhibits a universal linear trend, independent of centrality and \pt, indicating the initial shape is the sole contributor to the origin of the \PDz $v_2$. To further quantify the correlation, Fig.~\ref{fig:slope_intercept} presents the slopes (left plot) and intercepts (right plot) of the normalized \PDz $v_2$ versus normalized charged-particle $v_2$ as a function of \PDz \pt and centrality. The measured slopes are consistent with unity within uncertainties for all centrality bins across the entire \pt range. The intercepts, representing \PDz $v_2$ when charged-particle $v_2$ is zero, are consistent with zero. Combined with the slope measurements, these results strongly suggest that the \PDz elliptic flow originates primarily from the eccentricity of the initial geometry of the collision system~\cite{HIN-24-015_PAS}. Any nonlinearity observed in this correlation would indicate a potential contribution from other physical mechanisms to \PDz elliptic flow. Specifically, the possibility of a nonlinear correlation at high \PDz \pt could result from the pathlength-dependent energy loss of the charm quark in the QGP medium~\cite{PRC_D0_ch_v2}. 

\section{Summary}
The elliptic and triangular flow vectors ($v_2$ and $v_3$) of prompt and nonprompt \PDz in mid-rapidity ($\abs{y}<1$) in the \pt range 1--30\GeVc and three centrality classes in 0--50\% at $\sqrtsNN=5.02$~TeV are presented here. A strong \pt and centrality dependence is observed for prompt \PDz $v_2$, while the $v_3$ is essentially centrality independent. The $v_2$ of nonprompt \PDz shows a weak \pt dependence and a slight increase for more peripheral events, while a non-zero $v_3$ is observed with \pt 4--6\GeVc. The flow vectors of nonprompt \PDz are lower than those of prompt \PDz. A clear correlation between the initial collision geometry, characterized by the reduced flow vector, and the $v_2$ of prompt \PDz mesons in \PbPb collisions at $\sqrtsNN=5.02$~TeV is demonstrated. By employing event-shape engineering to systematically vary initial-state eccentricities, we have shown that the \PDz meson $v_2$ is strongly correlated to the eccentricity ($\epsilon_2$) of the initial geometry. This observation, coupled with the observed correlation between \PDz $v_2$ and charged-particle $v_2$ within the \pt range 1--3\GeVc, reinforces the conclusion that the initial geometry eccentricity is the dominant source of charm-hadron elliptic flow. This correlation is approximately linear and exists in the centrality region of 0--50\% (10\% classes) and for both low (2--10\GeVc) and high-\pt (10--30\GeVc) of \PDz-meson. A nonlinear correlation, specifically for high \PDz \pt, may be an indication of the path-length dependent energy-loss of the charm quark in the QGP medium. These findings provide compelling evidence for the significant role of initial-state geometry in shaping the collective flow of heavy quarks within the quark-gluon plasma created in heavy-ion collisions.

%

\end{document}